\RequirePackage{fix-cm}
\documentclass[smallextended]{svjour3}       
\smartqed  
\usepackage{graphicx}
\usepackage{amssymb}
\begin{document}

\title{Present time 
}

\titlerunning{Present time}        

\author{Gustavo E. Romero}


\institute{Instituto Argentino de Radioastronom{\'{i}}a (IAR, CCT La Plata, CONICET) \at
              C.C. No. 5, 1894, Villa Elisa, Buenos Aires, Argentina. \\
              Tel.: +54-221-482-4903\\
              Fax: +54-221-425-4909\\
              \email{romero@iar-conicet.gov.ar}
}

\date{Received: date / Accepted: date}

\maketitle

\begin{abstract}
The idea of a moving present or `now' seems to form part of our most basic beliefs about reality. Such a present, however, is not reflected in any of our theories of the physical world. I show in this article that presentism, the doctrine that only what is present exists, is in conflict with modern relativistic cosmology and recent advances in neurosciences.  I argue for a tenseless view of time, where what we call `the present' is just an emergent secondary quality arising from the interaction of perceiving self-conscious individuals with their environment. I maintain that there is no flow of time, but just an ordered system of events.

\end{abstract}
\vspace{0.3cm}

\begin{quotation}
\begin{flushright}
Again, the `now' which seems to bound the past and the future -- does it always remain one and the same or is it always other and other? It is hard to say.\\[0.5cm]
{\sl Aristotle.}
\end{flushright}
\end{quotation}  

\section{Introduction}
\label{intro}

Time has always puzzled philosophers and scientists alike. Traditionally, there are two broad views about the nature of time. These views are usually called the ``tensed'' and the ``tenseless'' views, or, for simplicity, the A and B theories of time. For an A-oriented person, only present things exist. There are many varieties of this ontological position: presentism, becoming theory, primitive tenses, branching universe theory, and so on. All of them distinguish the present in some way. In particular, presentism is the doctrine that it is always the case that, for every $x$, $x$ is present. The logical quantification in this definition is unrestricted, it ranges over all existents. In order to make this definition meaningful, the presentist must provide a specification of the term ``present''. A standard definition is:\\

{\bf Present}: The mereological sum of all objects with null temporal distance (Crisp 2003).\\

Since the mereological sum of objects is always an object, we can infer that for a presentist the present is an object, i.e. an individual with some properties. The open formula `$x$ is present' in the definition above, then means `$x$ is part of the mereological sum of all objects with null temporal distance'. It also purports that what there is exists only at an instant, not over a span of time.

A B-oriented person considers all this as pure nonsense. She maintains that past, present and future `equally' exist. For her, the fundamental temporal properties are relations of `earlier than', `later than' and `simultaneous with'. These are relations between events. There is no present in any absolute sense. The present is {\sl not} an object. Then, it cannot move, since only objects can move with respect to each other. There is no objective `flow' or passage of time.  

What is, then, the present in this view? My aim, in this article, is to answer this question from a B-perspective.


\section{Against Presentism}
\label{sec:ST}

The Englishman John McTaggart Ellis McTaggart presented a disproof of presentism in his famous paper {\sl Unreality of Time}  (McTaggart 1908). He reasoned as follows.
\begin{enumerate}
\item There is no time without change.
\item If time passes, events should change with respect to the properties of pastness, presentness, and futureness.
\item A given event, then, should be able to be in absolute sense, past, present and future. 
\item These properties exclude each other.
\end{enumerate}
Then: Events do not pass, just are. \\

There is no passage of time. There is no moving present. The mere idea of a flowing time simply does not make any sense. An additional problem is that if time flows, it should move with respect to something. If we say that there is a super-time with respect to which time flows, then we shall need a super-super-time for this super-time, and we shall have an infinite regress. In addition, there is no flow without a rate of flow. At what rate does time go by? The answer 1 sec per sec is meaningless. It is like saying that a road extends along a distance  of one km per each km that it extends! 

On the physical side, the theory of special relativity seems not to be friendly to the idea of an absolute present, at least in its usual Minkowskian 4-dimensional interpretation. Special relativity is the theory of moving bodies formulated by Albert Einstein in 1905 (Einstein 1905). It postulates the Lorentz-invariance of all physical law statements that hold in a special type of reference systems, called {\it inertial frames}. Hence the `restricted' or `special' character of the theory. The equations of Maxwell electrodynamics are Lorentz-invariant, but those of classical mechanics are not. When classical mechanics is revised to accommodate invariance under Lorentz transformations between inertial reference frames, several modifications appear. The most notorious is the impossibility of defining an absolute simultaneity relation between events. Simultaneity results to be frame-dependent. Then, some events can be future events in some reference system, and present or past in another system. Since what exists cannot depend on the reference frame adopted for the description of nature, it is concluded  that past, present, and future events exist. Consequently, presentism, the doctrine that only what is present exists, is false.

The presentist or A-theorist of time might find a way around this argument adopting a different (purely Lorentzian) interpretation of the theory (Crisp 2008, Zimmerman 2011), which relinquishes the concept of space-time. The problems of this approach have been discussed at length by Saunders (2002), and I do not insist on the topic here. Instead, I prefer to say a few words about the much less discussed issue of the compatibility of presentism with general relativity. 

General relativity is the theory of space, time and gravitation proposed by Einstein in 1915 (Einstein 1915). Space-time is an indispensable ingredient of this theory. A 4-dimensional real and differentiable manifold is adopted to represent space-time, the physical aggregation of all events (Romero 2013). A rank-2 metric field $g_{ab}$ is defined over the manifold to represent the potentials of the gravitational field. Distances over the manifold are given by
\begin{equation}
ds^2=g_{ab} dx^a dx^b,
\end{equation}
where $dx^a$ is a 4-dimensional differential length vector. The key issue to determine the geometric structure of space-time, and hence to specify the effects of gravity, is to find the law that fixes the metric once the source of the gravitational field is given. The source of the gravitational field is matter. The energy-momentum tensor $T_{ab}$  represents the physical properties of material things that generate space-time. The curvature of space-time at any event is related to the energy-momentum content at that event by a set of differential equations. 

These equations, the Einstein field equations, can be written in the simple form:

\begin{equation}
	G_{ab} = -\frac{8\pi G}{c^4}\, T_{ab}, 
	\label{einstein}
\end{equation}
where $G_{ab}$ is the so-called Einstein's tensor, which is linear in the curvature\footnote{The curvature is represented by the Riemann tensor $R_{abcd}$, formed with second derivatives of the metric (see, e.g. Hawking and Ellis 1973).} and non-linear in the metric. It contains all the geometric information on space-time. The constants $G$ and $c$ are the gravitational constant and the speed of light in vacuum. Einstein's field equations are a set of ten non-linear partial differential equations for the metric coefficients. 

A crucial point of general relativity is that the 4-dimensional space-time with non-zero curvature is not dispensable anymore. Contrarily to the Minkowskian case, general relativity is not susceptible of a global Lorentzian formulation. This poses a problem for presentism, because of the relativity of simultaneity implied by the constancy of the speed of light in space-time. However, for some cosmological models ${\cal M}_{\rm st}=<M,\;g_{ab},\;T_{ab}>$ a kind of `cosmic time' can be re-introduced in space-time, and some presentists have tried to use it to their advantage.

Thomas Crisp (2008) has proposed a ``presentist-friendly'' model of general relativity. He suggests that the world is represented by a 3-dimensional space-like hypersurface that evolves in a forth dimension (time). This interpretation requires the introduction of a preferred foliation of space-time at large scales, and to consider the $3+1$ usual decomposition for the dynamics of space-time in such a way that `the present' is identified with the evolving hypersurface. This situation is depicted in Figure 1. 

\begin{figure} 
\centering
\includegraphics[width=25pc]{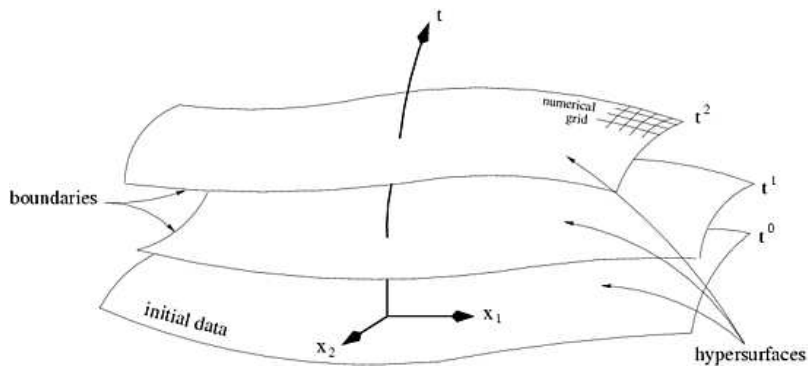}
\caption[]{A `presentist-friendly' space-time: Evolving 3-dimensional space-like surfaces in a space-time with a preferred time-direction.}
\label{Fig1}
\end{figure}   


In order to set up such a model for space-time, some global constraints must be imposed: there should be a unique foliation into surfaces of constant mean curvature. This is the case, for instance, of the Friedmann-Robertson-Walker-Lema\^itre metric. This metric is isotropic and homogeneous. General conditions for such kind of metrics require the absence of Cauchy horizons, the fulfillment of the so-called {\it energy conditions} (Hawking and Ellis 1973), and symmetry constraints. In this kind of metrics, the parameter along which the hypersurfaces evolve is called `cosmic time'. 

I confess that I do not see how such  `cosmic time' can help the presentist's cause. The foliation of a manifold is nothing else than a computational device. The selection of a given hypersuface as `present' is completely arbitrary. A hypersurface is nothing else than a class of events (i.e. a concept, not a thing), which we decide to specify as initial data for subsequent calculations. Any hypersurface can be used for this purpose, and since the Einstein equations are time-reversible, we can compute the evolution with respect to both $t$ and $-t$. There is not reason for thinking that a particular class of events, even if they are all space-like, is the `present'. Moreover, classes do not flow, as the present is supposed to do.

A similar criticism is valid for the definition of present given by Crisp (2003), who considers that the present is the aggregation of all things (I would say events) with null temporal distance.  There are events with null temporal distance in the past of the event of reading this line, which are simultaneous with  the landing of Apollo 11 on the moon on July 20th 1969. There is no reason to think that your reading of this line is `now' and not the events simultaneous with Armstrong's remarkable step: both events form part of aggregates with null temporal distances. Why one aggregate of events is present and the other is not? 

There are additional problems related to the actual structure of the universe if it is supposed to be represented by a smoothly foliable theoretical space-time model ${\cal M}_{\rm st}$. First, there are black holes. Non-spherically symmetric black holes (i.e. rotating or Kerr black holes) have Cauchy horizons. There is no way to compute the evolution of any physical system inside these regions, whatever be the external space-time foliation. There is no possibility of synchronization between clocks outside and inside the inner horizon of a Kerr black hole. And the evidence for rotating black holes in the universe is overwhelming (e.g. Romero and Vila 2013). This point cannot be ignored by the presentist.

Another problem for the `presentist-friendly' space-time is that observational data indicate that remote supernovae Type Ia present redshifts that either suggest the universe is expanding in an accelerated way or it is inhomogeneous. Both possibilities are ruinous for the presentist. The first requires massive violations of the energy conditions in the universe. These violations can be produced by dark fields with negative energy densities or by gravity with modified dynamical equations. In either case, particle cosmological horizons appear in the universe, disconnecting different regions and making global time synchronization impossible. On the other hand, if the universe is inhomogeneous or anisotropic on medium scales, then the  Friedmann-Robertson-Walker-Lema\^itre model is not a correct description of space-time and more complex models should be considered. No foliation of constant curvature is possible with inhomogeneity or anisotropy, with the consequent problem for synchronization. 

Zimmerman (2011) has pointed out that in a desperate case of conflict with general relativity the presentist can abandon Einstein's theory, since one can be sure that relativity is ultimately a wrong theory because it is incompatible with quantum mechanics. I protest. General relativity is {\it not} inconsistent with quantum mechanics as it is sometimes loosely stated. The background space-time of quantum mechanics is flat Minkowskian space-time. Even in a space-time with non-zero curvature, quantum mechanical calculations can be performed (Wald 1994). General relativity is a classical theory, and hence it cannot deal with {\it quantum interactions of the gravitational field}. This is something very different from saying the there is incompatibility with quantum mechanics or quantum field theory. What is not known is what a quantum field theory {\it of gravitation} is. What we actually know is that at the scales that are relevant for the presentist, general relativity is a well-tested theory. But even if it is replaced by other field theory to better accommodate the phenomenology of dark matter and the apparent universal accelerated expansion, the very same problems I mentioned above will remain.  For the presentist, the battle is lost from the beginning: the very concept of space-time is at odds with presentism. And this is because space-time is the ontological sum of all events.  The mere postulation of space-time implies a consent to events that can be classified as past or future with respect to some other events.  Space-time is inconsistent with presentism.  
  
Said all that, yet, we all have a kind of feeling that ``our time is running out''. Where does this feeling come from? To answer we should look not at space-time, but into our own brains. 

\section{When is `now'?}
\label{sec:2}

If the present is an instant of time instead of a thing, then the question of ``which instant is present?'' follows. One possible answer is ``now''. But...when is `now'?

`Now', like `here', is an indexical word. To say that I exist now gives no information on when I exist. Similarly, to say that I am here, gives no information on where I am. There is no particular moment of time defined as an absolute now. 

I maintain that `nowness' and `hereness' emerge from the existence of perceiving self-conscious beings in a certain environment. What these beings perceive is {\sl not} time, but changes in things, i.e. events (Bunge 1977). Similarly, they do not perceive space, but spatial relations among things. In particular, we do not perceive the passage of time. We perceive how our brain changes. I claim that there is no present {\sl per se}, in the same way that there is no smell, no pain, no joy, no beauty, no noise, no secondary qualities at all without sentient beings. What we call ``the present'' is not in the world. It emerges from our interaction with the world. 

We group various experienced inputs together as present; we are tempted to think that this grouping is done by the world, not by us. But this is just delusional.  
I maintain that tenses are not needed and in fact are not wanted by the natural sciences. This idea is clearly expressed by E. Poeppel on the basis of neurological research (Poeppel 1978):

\begin{quotation}

[...] our brain furnishes an integrative mechanism that shapes sequences of events to unitary forms...that which is integrated is the unique content of consciousness which seems to us present. The integration, which itself objectively extends over time, is thus the basis of our experiencing a thing as present.

[...] The now, the subjective present, is nothing independently; rather it is an attribute of the content of consciousness. Every object of consciousness is necessarily always now - hence the feeling of nowness.
\end{quotation}

The perception of motion gives an additional argument against the idea that the present is an instant of time. According to Le Poidevin (2009): 

\begin{enumerate}

\item What we perceive, we perceive as present.
\item We perceive motion.
\item Motion occurs over an interval.

\end{enumerate}

Therefore: What we perceive as present occurs over an interval. \\

Recent research in neurosciences lends strong support to these claims. Perception of events outside the brain and the construction of what we call time is a complex cluster of processes that involves different cortical and sub-cortical regions. Distortions in timing can be produced by narcotics, experimental manipulation, strong emotions, and by different brain disorders such as Alzheimer's disease, clearly indicating a dependence of temporal experience on brain processes. The involvement of sub-cortical areas in external change perception explains why extreme fear and other abnormal emotional states can modify the subjective experience of time (e.g. Stetson et al. 2007). 

A very important breakthrough in neurological research about the timing mechanisms operating in the brain was made by Benjamin Libet and collaborators (Libet et al. 1964, Libet 1973). In a series of now classical experiments, Libet et al. demonstrated that there is a time delay of about 0.5 s between the starting of brain stimulation and the appearance of awareness of the stimulus. This shows that awareness of an event happens in the brain when the event is past: what we become aware of has already occurred about 0.5 s earlier. In Libet's words: ``We are not conscious of the actual moment of the present. We are always a little late." (Libet 2004). The entire battery of sensory stimuli are manipulated by the brain to create a coherent representation of the external world in such a way that we are not aware of any time delay. The subjective `present' is actually a construction made with a manifold of sensory information of events {\em in the past}.

The motor system does not wait $\sim0.5$ s before making its decisions. These are done unconsciously and over spans as short as 10 ms in some cases. Consciousness allows further interpretation and adjustments on the basis of later information (Eagleman et al. 2000). The actual span required to create a transient representation of the environment can vary from an individual to another, but should take more than 100 ms on average. In Eagelman's words (Eagleman 2009):

\begin{quotation}
This hypothesis --that the system waits to collect information over the window of time during which it streams in--
applies not only to vision but more generally to all the other senses. Whereas we have measured a tenth-of-a-second
window of postdiction in vision, the breadth of this window may be different for hearing or touch. If I touch your toe
and your nose at the same time, you will feel those touches as simultaneous. This is surprising, because the signal from
your nose reaches your brain well before the signal from your toe. Why didn't you feel the nose-touch when it first
arrived? Did your brain wait to see what else might be coming up in the pipeline of the spinal cord until it was sure it had
waited long enough for the slower signal from the toe? Strange as that sounds, it may be correct.

It may be that a unified polysensory perception of the world has to wait for the slowest overall information. Given
conduction times along limbs, this leads to the bizarre but testable suggestion that tall people may live further in the past
than short people. The consequence of waiting for temporally spread signals is that perception becomes something like
the airing of a live television show. Such shows are not truly live but are delayed by a small window of time, in case
editing becomes necessary.
\end{quotation}    

All evidence from neuroscience research points to the hypothesis the `the present' is a construction of the brain; a construction that is not instantaneous. We do not perceive time; we only are aware of events and can compare the event rate or their clustering in the external world with the rate of activity of our own brain (e.g. Karmarkar and Buonano 2007).

Any tentative definition of `present' compatible with modern neuro-biology science must take into account the role of the perceiving and sentient individual. In the next section I offer some provisional definitions that meet this requirement and distinguish among the different meanings in which the word `present' is used.

\section{Defining the present}
\label{sec:3}
 
Physical events are ordered by the relations `earlier than' or `later than', and `simultaneous with' (Gr\"unbaum 1973). There is no `now' or `present' in the mathematical representation of the physical laws. 
What we call `present' is not an intrinsic property of the events nor an instant of time, much less a moving thing. `Present' is a concept abstracted from the relation between a certain number of events and a self-conscious individual.\\

{\bf Present}: Class of all events simultaneous with a given brain state.\\

To every brain state there is a corresponding present. The individual, notwithstanding, needs not to be aware of all events that form the present. The present, being a class of events, is an abstract object without any causal power.\\

{\bf Psychological present}: Class of local events that are causally\footnote{For a complete account of causality as a relation between events see Bunge (1979).} connected to a given brain state. \\

Notice that from a biological point of view only local events are relevant. These events are those that directly trigger neuro-chemical reactions in the brain. Such events are located in the immediate causal past of the brain events that define the corresponding state. The psychological present is a conceptual construction of the brain, based on abstraction from events belonging to an equivalence class. The present, then again, is not a thing nor a change in a thing (an event). It is a construction of the brain; a fiction albeit a very useful one for survival. Yet again, individuals are not necessarily aware of {\it all} events that are causally relevant for the construction of the psychological present.

E.R. Kelly (1882) introduced the concept of `specious present', which William James elaborated  as ``the short duration of which we are immediately and incessantly sensible'' (James 1908). I propose to update this definition to:\\ 

{\bf Specious present}: Length of the time-history of brain processes necessary to integrate all local events that are physically (causally) related to a given brain state.\\

The specious present, being related to brain processes, can be different for different individuals equipped with different brains. The integration of the specious present can be performed in different ways, depending on the structure of the brain. It is even possible to imagine integration systems that can produce more than one specious present or even systems that might `recall' the future (see Hartle 2005 for examples based on computers). If biological evolution has not produced such systems, it seems because of the existence of space-time asymmetric boundary conditions that introduce a preferred direction for the occurrence of processes (Romero and P\'erez 2011).

Finally, I introduce a {\sl physical present}.\\

{\bf Physical present}: Class of events that belong to a space-like hypersurface in a smooth and continuous foliation of a time-orientable space-time. \\

Since in the manifold model of space-time every event is represented by an element of the manifold, the introduction of this class does not signal a special time identified with `now'. Every space-like hypersurface corresponds to a different time and none of them is an absolute present `moving' into the future. Actually, naming `the future' to a set of surfaces in the direction opposite to the so-called Bing Bang is purely conventional.

\section{Final remarks}

I have distinguished three different types of present: psychological, physical, and specious. The former two are classes of events, hence they are concepts. The latter is not an instant of time but an interval in space-time associated with the world history of a sentient individual.

In any case, the present does not flow or move. Only material individuals (and their brains, if they have one) can change and move. Becoming is not a property of physical events, but of the consciousness  of the events. We call `becoming' to the series of states of consciousness associated with a certain string of physical changes. Events do not become. Events just {\em are}.

\section*{Appendix: Present time in quantum mechanics}
\label{Appendix}

Quantum mechanics has been invoked to justify almost every conceivable belief, from free will to the existence of God. It is not surprising that the presentists have also appealed to it. Two main arguments have been proposed to support presentism: the so-called `collapse of the wave function', and the non-locality implied by the experimental violation of Bell's inequalities. In this appendix I discuss and dismiss both arguments.  

The collapse of the wave function is supposed to involve absolute simultaneity.  Absolute simultaneity, in turn, allows to classify events into past, present and future {\it \'a la} Newton. The argument runs as follows, in words of Michael Tooley (Tooley 2008):

\begin{quotation}
Consider an electron that has been fired towards the screen of a cathode ray tube. According to quantum mechanics, the initial conditions do not causally determine what point on the screen the electron will strike. All that follows from the laws of quantum mechanics is that {\it there are various non-zero probabilities} of the electron's striking different parts of the screen. But once the electron strikes the screen, {\it the probabilities of hitting other parts of the screen must change} from having non-zero values to being equal to zero. But then must it not be the case that that all these changes are absolutely simultaneous with the event that consists of the electron's hitting the screen a certain location?\footnote{My italics.} 
\end{quotation}

This is abhorrent. It is claimed that probabilities, a mathematical construct that represents a certain propensity of a physical system, can change in space and time. This reification of the concept of probability is wrong independently of quantum mechanics. Think, for instance, of a roulette. A croupier spins a wheel in one direction, then spins a ball in the opposite direction around the tilted circular track running around the circumference of the wheel. The ball eventually loses momentum and falls onto the wheel and into one of 37 (in French/European roulette) or 38 (in American roulette) colored and numbered pockets on the wheel. Let us consider a French roulette. There is an a priori probability of 1/37 for a given number to be the winner. After the ball has fallen into a given pocket, say number 7, the probability for this number is still 1/37, no 1!  Because of this low probability is that the casino is paying you. Similarly, if you toss a coin, there is a probability of 1/2 of getting heads both before and after a flipping. Probabilities are not a physical field that suddenly collapses at some moment of time. As all mathematical constructs, probabilities are timeless. 

Even the so-called `collapse of the wave function' is non-sense since the wave function is a mathematical function defined in a Hilbert space, not an entity propagating through space-time, and hence it cannot collapse at all. The wave function allows to estimate probabilities of a given event. It is a solution of the Schr\"odinger or Dirac equations, which are linear differential equations. The physics of the non-linear interaction of the quantum system represented by the wave function and the environment is surely non-linear and must be discussed in the context of a theory of quantum measurement, that depends on the details of the detector. Certainly, it is not enough just to add a postulate (as von Neumann's) to account for such a complex theory (see Bunge 1967, Perez-Bergliaffa et al. 1993). 

Let us turn to the claim that quantum non-locality reestablishes absolute simultaneity and introduces a preferred foliation of space-time. Einstein, Podolsky and Rosen (1935)  -EPR- were the first to notice that quantum mechanics might imply non-locality. Consider two electrons prepared in a singlet state and fired in opposite directions from a central source with equal velocities. This event occurs at a time $t_0$.  Quantum mechanics describes the system with a single two-particle wave function that is not the product of independent particle wave functions. Because electrons are indistinguishable particles, it is not proper to say that electron 1 goes this way and electron 2 that way.  At a later time $t_1>t_0$, a measurement of one electron's momentum\footnote{Although I am considering the momentum, a conserved quantity, in this example, conservation laws do not seem to play a role in the phenomenon, since these kind of correlations are observed in the polarization of photons, which certainly in not a conserved quantity (see Maudlin 2002).} would instantly reveal the momentum of the other electron - without need of measuring it. This is a non-local correlation that seems to violate relativity if the separation of the electrons is space-like. Schr\"odinger described the two electrons as ``entangled" (verschr\"ankt) at their first measurement, so the EPR  ``non-locality" phenomenon is also known as ``quantum entanglement".

Contrary to Einstein's expectations, experiments have generally favored quantum mechanics as a description of nature, over local hidden variable theories. Any physical theory that supersedes or replaces quantum theory must make similar experimental predictions and must therefore also be non-local in the specific sense of showing the existence of space-like correlations for entangled quantum systems. Does this imply the existence of the present? I say no.

The existence of correlations between components of entangled systems does not require
superluminal matter transport or signaling. Much less of information. Information is a
property of languages, not of physical systems. Contrary to what Tim Maudlin (2002) states, superluminal transmission of information is not required by quantum entanglement.
Information has not, and cannot, have any effect upon physical systems.

The correlations just show that once a system is formed, it remains a system. Although the actual mechanism that enforces the systemic memory is not clear, quantum non-locality cannot be used to define an absolute simultaneity, since far from be universal, its very nature requires it to be selective and operating only on systems that were prepared in a singlet state once. There is no way to know the output of a measurement on the second electron of our example above if the data on the preparation and the first measurement were not transmitted to the second detector. This transmission is only possible at the speed of light, at most.    

I conclude that the concept of present is alien to quantum mechanics.



\bibliographystyle{aipproc}   


\newpage

\section*{Gustavo E. Romero} Full Professor of Relativistic Astrophysics at the University of La Plata and Chief Researcher of the National Research Council of Argentina. A former President of the Argentine Astronomical Society, he has published more than 300 papers on astrophysics, gravitation and the foundations of physics. Dr. Romero has authored or edited 9 books (including {\sl Introduction to Black Hole Astrophysics}, with G.S. Vila, Springer, 2013). His main current interest is on black hole physics and ontological problems of space-time theories.

\end{document}